\documentclass[apj,useAMS,usenatbib,usegraphicx,upmath]{emulateapj} \usepackage{amsmath} \usepackage{amsfonts} \usepackage{amssymb} \usepackage{apjfonts} \usepackage{color}

\newcommand{\rd}{\mathrm{d}} 
\newcommand{\rD}{\mathrm{D}} 
\newcommand{\BV}{Brunt--V\"ais\"al\"a} 
\newcommand{\TG}{\textit{Taylor--Goldstein} } 
\newcommand{\HD}{HD\,209458\,b} 
\newcommand{\upartial}{\partial} 
\renewcommand{\vec}{\mathbf} 

\newcommand{\cdel}[1]{} 

\def\grad{\mbox{\boldmath{$\nabla$}}}
\def\mathbi#1{\textbf{\em #1}}

\shorttitle{GRAVITY WAVES ON HOT EXTRASOLAR PLANETS I} 
\shortauthors{WATKINS \& CHO}
\submitted{}

\begin{document}

\title{Gravity Waves on Hot Extrasolar Planets:\\
  I.  Propagation and Interaction with the Background}

\author{Chris Watkins and James Y-K. Cho\altaffilmark{1}}

\affil{Astronomy Unit, School of Mathematical Sciences, Queen Mary,
  University of London, Mile End Road, London E1 4NS, UK}

\email{c.watkins@qmul.ac.uk; J.Cho@qmul.ac.uk}

\altaffiltext{1}{Visiting scientist, Department of Terrestrial Magnetism,   Carnegie Institution of Washington, Washington, DC 20015, USA}

\begin{abstract}
  We study the effects of gravity waves, or $g$-modes, on hot extrasolar planets.  These planets are expected to possess stably-stratified atmospheres, which support gravity waves.  In this paper, we review the derivation of the equation that governs the linear dynamics of gravity waves and describe its application to a hot extrasolar planet, using \HD\ as a generic example.  We find that gravity waves can exhibit a wide range of behaviors, even for a single atmospheric profile.  The waves can significantly accelerate or decelerate the background mean flow, depending on the difference between the wave phase and mean flow speeds.  In addition, the waves can provide significant heating ($\sim\!  10^2$ to $\sim\! 10^3$~K per planetary rotation), especially to the region of the atmosphere above about 10 scale heights from the excitation region.  Furthermore, by propagating horizontally, gravity waves provide a mechanism for transporting momentum and heat from the dayside of a tidally locked planet to its nightside.  We discuss work that needs to be undertaken to incorporate these effects in current atmosphere models of extrasolar planets.  \end{abstract}


\section{Introduction}

A stably-stratified atmosphere, characterized by a positive vertical entropy gradient, can support gravity waves, or $g$-modes.  Gravity waves are oscillations which arise from the buoyancy of parcels in the fluid. Such waves are readily excited by flow over thermal and surface topography, convective and shear instabilities, and flow adjustment processes.  They propagate through the atmosphere both horizontally and vertically.  Gravity waves in the terrestrial atmosphere and ocean are much studied \citep[e.g.,][]{Gossard1975,Gill1982}.  They have also been observed on other Solar System bodies, such as Jupiter \citep{Young1997} and Venus \citep{Apt1980}.

In the terrestrial atmosphere, a typical gravity wave has an energy flux of approximately 10$^{-3}$ to 10$^{-1}$~W~m$^{-2}$.  Despite being small, compared to the total amount of absorbed solar flux ($\sim$237~W~m$^{-2}$), gravity waves are responsible for significantly modifying---even dictating---large-scale flow and temperature structures.  Several well known examples of this are the Quasi-Biennial Oscillation, reversal of mean meridional temperature gradient in the upper middle atmosphere, and generation of turbulence \citep[e.g.,][]{Andrews1987}.  We expect similar effects to be present in the atmospheres (and oceans) of extrasolar planets.  Moreover, due to the greater irradiation and scale heights on them, the acceleration and heating effects of gravity waves can be much stronger on hot extrasolar planets.

There has been much interest in modeling the atmospheric circulation of extrasolar planets \citep[e.g.,][]{Joshi1997,Showman2002,Cho2003,Cho2008a,Burkert2005,Cooper2005,Dobbs-Dixon2008,Koskinen2007,Langton2007,Langton2008,Menou2009,Showman2008}.  Accurate simulation of atmospheric circulation is crucial for interpreting observations of extrasolar planets, as well as for improving theoretical understanding in general.  For this, the role of eddies and waves in transferring momentum and heat needs to be addressed \citep{Cho2008b}. This has long been recognized in Solar System planet studies \citep[e.g.,][]{Lindzen1990,Fritts2003}.

The plan of the paper is as follows.  In \S\ref{sec:theory} we derive the governing equation appropriate for linear monochromatic gravity waves on hot extrasolar planets. We also discuss a simple parameterization of the key non-linear process, saturation.  In addition, we present solutions to the equation for simple isothermal atmospheres, with and without shear in the background mean flow.  In \S\ref{sec:application} we extend the calculation to a physically more realistic situation, by using background flow and temperature profiles derived from a three-dimensional (3-D) hot--Jupiter atmospheric circulation simulation.  This is the first such calculation to have been performed for extrasolar planets.  Through this, the significant effects of gravity waves on hot extrasolar planet atmospheric mean flows are demonstrated.  In this section, we also discuss a way in which gravity waves can transport momentum and heat horizontally---e.g., from the dayside to nightside on tidally locked planets.  In \S\ref{sec:implications} we discuss the implications of our work for current extrasolar planet atmospheric modeling work.  We conclude in \S\ref{sec:conclusion}.

\section{Linear Theory}\label{sec:theory}

\subsection{\TG Equation}\label{tge} 

The dynamics of a linear gravity wave is described by the {\it   Taylor--Goldstein equation} (TGE).  This equation is derived from the full, 3-D hydrodynamics equations \citep{Batchelor1967}. In this work, we restrict the description to two dimensions and neglect rotation:
\begin{subequations}\label{hydroeqn}
\begin{eqnarray}
  \frac{\rD \vec{u}}{\rD t}\ & = &\ -\frac{1}{\rho}\grad p + 
  \vec{g}   \\ 
  \frac{\rD \rho}{\rD t}\ & = &\ -\rho\grad \cdot \vec{u} \\
  \frac{\rD \theta}{\rD t}\ & = &\ \frac{\theta}{c_p T}\dot{Q}\, , \end{eqnarray}
\end{subequations}
where $\rD / \rD t = \upartial / \upartial t + \vec{u} \cdot \grad$; $\vec{u} = (u,w)$ is the flow in the horizontal and vertical directions $(x,z)$, respectively; $\rho$ is the density; $\vec{\grad} = (\upartial / \upartial x, \upartial / \upartial z$); $p$ is the pressure; $\vec{g} = (0,-g)$ is the gravity; $\theta$ is the potential temperature; $c_p$ is the specific heat at constant pressure; $T$ is the sensible temperature; and $\dot{Q}$ is the net diabatic heating rate.  Equations~(\ref{hydroeqn}) are supplemented with the ideal gas law:
\begin{equation} 
  p = \rho R T\, ,
\end{equation}
where $R$ is the specific gas constant.  Note that
\begin{equation} 
  \theta \equiv 
  T \left(\frac{p_{\scriptscriptstyle R}}{p}\right)^{\kappa},
\end{equation}
where $p_{\scriptscriptstyle R}$ is some reference pressure (here taken to be the pressure at the lower boundary of the model) and $\kappa = R/c_p$; $\theta$ is related to the entropy $s$ by $\rd s = c_p\,\rd\ln\theta$.

The neglect of the third dimension and rotation requires that we restrict our analysis to waves with horizontal scale $L \la U/\Omega$, where $U$ is the characteristic mean flow speed and $\Omega$ is the planetary rotation rate.  This scale is adequate for all gravity waves, except for large-scale tides (which has been recently considered by \citet{Gu2009}).  As an example, $U/\Omega \sim 10^7$~m for \HD, based on $U$ in hot extrasolar planet simulations of \citet{Thrastarson2009}.  The resulting value is approximately 1/10 of the planet's radius.  In addition, $\vec{g}$, $R$, and $c_p$ are taken to be constant and $\dot{Q}$ is specified.  These restrictions do not mitigate the basic application and implications presented in this work.  However, for broader applications, relaxation of these and other restrictions will be considered in future work.

The variables in equations~(\ref{hydroeqn}) are all expanded as a small perturbation about a mean value, which is a function of height only: 
\begin{equation}\label{expan} 
  \zeta(x,z,t)\ =\ \zeta_0(z) +   \zeta_1(x,z,t)\, .
\end{equation}
For the thermodynamic variables, we require that $\zeta_1 / \zeta_0 \ll~1$; however, this is not required for the flow variables, $u$ and $w$.  The mean state is assumed to be in hydrostatic balance, $\rd p_0/\rd z = -\rho_0\, g$, and contains only horizontal flow so that $w_0 = 0$.  We also assume the anelastic approximation \citep[e.g.,][]{Ogura1962}, $\grad\cdot(\rho_0\vec{u}) = 0$.  This implies the following:
\begin{subequations}\label{anelas}
  \begin{eqnarray}
    \frac{N^2 H^2_{\rho}}{c^2_s}\ & \ll &\ 1 \\ 
    \frac{\gamma D}{H_p}\ & \la &\ 1 \\
    \left| \frac{u_0}{w_1} \right| \frac{D}{L}\ & \la &\ 1, 
  \end{eqnarray}
\end{subequations}
where $N(z) = [g\,(\rd \ln \theta_0 / \rd z)]^{1/2}$ is the \BV\ frequency; $H_{\rho}(z) \equiv |\rho_0\,(\rd \rho_0/\rd z)^{-1}|$ and $H_p(z) \equiv |p_0\,(\rd p_0/\rd z)^{-1}|$ are the density and pressure scale heights, respectively; $\gamma =~c_p/c_v$ is the ratio of specific heats, with $c_v$ the specific heat at constant volume; $c_s=(\gamma RT)^{1/2}$ is the speed of sound; and, $D$ is the vertical scale of the motion.

With (\ref{expan}) and (\ref{anelas}), we obtain
\begin{subequations}\label{linear}
\begin{eqnarray} 
  \frac{\upartial u_1}{\upartial t} + 
  u_0 \frac{\upartial u_1}{\upartial x} + 
  w_1 \frac{\rd u_0}{\rd z}\ & = & \ 
  -\frac{\upartial  \Phi_1}{\upartial x} \\
  \frac{\upartial w_1}{\upartial t} + 
  u_0 \frac{\upartial w_1}{\upartial x}\ & = &\  
  -\frac{\upartial\Phi_1}{\upartial z} + g \Theta_1 \\ 
  \rho_0 \frac{\upartial u_1}{\upartial x} + 
  \rho_0\frac{\upartial w_1}{\upartial z} + 
  w_1\frac{\rd\rho_0}{\rd z}\ & = &\  0 \\
  \frac{\upartial \Theta_1}{\upartial t} + 
  u_0\frac{\upartial \Theta_1}{\upartial x} + 
  w_1\frac{\rd \ln\theta_0}{\rd z}\ & = & \ 
  \frac{\dot{Q}}{c_p T_0}\, ,
\end{eqnarray}
\end{subequations}
where $\Phi_1 = p_1 / \rho_0$ and $\Theta_1 = \theta_1 / \theta_0$. Since the coefficients in equation~(\ref{linear}) are independent of $x$ and $t$, we assume perturbations of the form,
\begin{equation}
  \zeta_1(x,z,t)\ =\ \tilde{\zeta}(z)\,\exp\{i(kx - \omega t)\}\, ,
\end{equation}
where it is understood that the real part is to be taken.  This leads to the polarization equations:
\begin{subequations} 
\begin{eqnarray}
  -ik \left(c-u_0 \right) \tilde{u} + \frac{\rd u_0}{\rd z} \tilde{w}\   
  & = &\ -ik \tilde{\Phi}\\
  -ik \left(c-u_0 \right) \tilde{w}\ & = &\ 
  -\frac{\rd \tilde{\Phi}}{\rd z} + g \tilde{\Theta} \\
  -ik \tilde{u}\ & = &\ \frac{\rd \tilde{w}}{\rd z} -             \frac{\tilde{w}}{H_{\rho}} \label{eq:u1polar} \\
  -ik \left(c-u_0 \right) \tilde{\Theta} + \frac{N^2}{g} \tilde{w}\ 
  & = &\ \tilde{F},
\end{eqnarray}
\end{subequations}
where $F \equiv \dot{Q}/(c_pT_0)$ is the forcing, $c = \omega/k$ is the constant (possibly complex, see \S\ref{sec:saturation}) horizontal phase speed, and $(c - u_0)$ is the intrinsic phase speed.  Now, transforming to a new variable so that the effect of decreasing density with height is compensated,
\begin{subequations}\label{eq:newvar}
\begin{eqnarray} 
  \hat{w}(z)\ & = &\ \tilde{w}\, \exp\{ -\chi(z) \} \\ 
  \chi(z)\ & = &\ \int^z_{z_b} \frac{\rd \xi}{2H_{\rho}(\xi)}\, ,   
\end{eqnarray} 
\end{subequations}
where $z_{\rm b}$ is $z$ at the bottom, we obtain the TGE:
\begin{equation}\label{eq:TGE}  
  \frac{\rd^2 \hat{w}}{\rd z^2} + m^2 \hat{w}\ =\     \frac{\kappa\,\dot{Q}}{H_p \left(c-u_0\right)^2}\, e^{-\chi}\, .
\end{equation}
Here, $m = m(z)$ is the index of refraction, and corresponds to the local vertical wavenumber. It is given by
\begin{equation}\label{eq:indref} 
  m(z)\ =\ \left[\frac{N^2}{\left(c-u_0\right)^2} +                       \frac{u_0''}{\left(c-u_0\right)} +                           \frac{u_0'}{H_{\rho}\left(c-u_0\right)} - \frac{1}{4H^2_{\rho}} -         k^2 \right]^{1/2}. 
\end{equation} 
In equation~(\ref{eq:indref}) we have used $H_p = RT_0(z)/g$ and the prime indicates differentiation with respect to $z$. 

The vertical structure of the perturbations, oscillating in $z$, signify that we are dealing with {\it internal} waves.  In equation~(\ref{eq:indref}), the key terms contributing to $m^2$ are the first (``buoyancy'') term and the last (``non-hydrostatic'') term. Although the other three terms contribute, generally the buoyancy and non-hydrostatic terms control whether the wave propagates since the flow shear and curvature are small and the scale height is large in practice.  For large wavelength waves, the waves are hydrostatic and the non-hydrostatic term is small; then, the buoyancy term dominates. In these cases, as long as the atmosphere is stratified (i.e., $N^2 > 0$) and $c \neq u_0$, the wave will propagate vertically. However, for shorter, non-hydrostatic waves, it is possible that $k^2 > [N / (c - u_0)]^2$. In these cases, $m$ is imaginary and the wave does not propagate. This situation is discussed in more detail in \S\ref{subsec:WBI}.

\begin{figure}
  \epsscale{1.08}
  \plotone{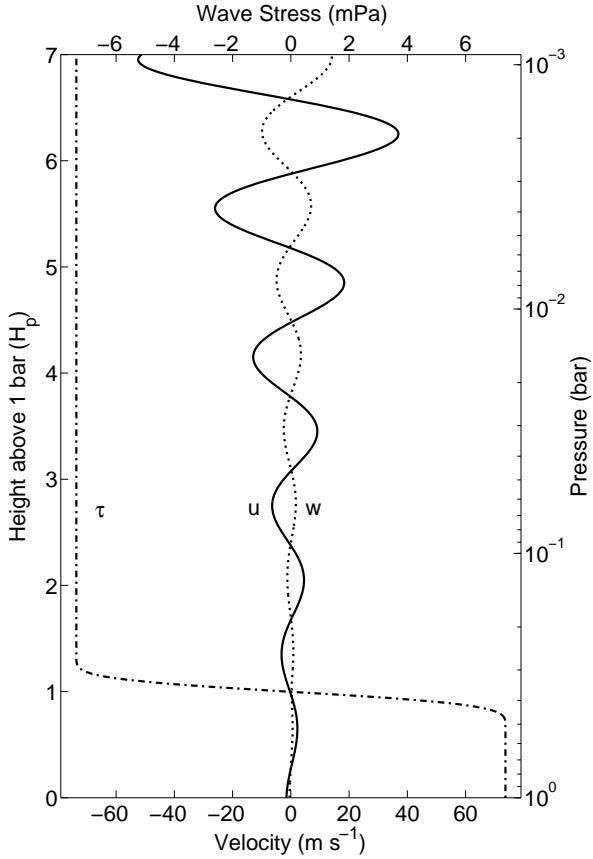}
  \caption{A gravity wave propagating in an isothermal ($T_0 = 1350$~K) and     constant background flow ($u_0 = 350$~m~s$^{-1}$) atmosphere.  The phase     speed of the wave $c$ is 100~m~s$^{-1}$ and the horizontal wavelength,     $2\pi / k$, is 2500~km.  The vertical perturbation velocity $w$     ($\cdots$), horizontal perturbation velocity $u$~(---), and wave stress     $\tau$ ($-\! \cdot\!  -$) are shown.  The latter is the vertical transport     of the horizontal momentum.  Wave amplitudes, $u$ and $w$, grow     exponentially with height, but the wave stress is constant with height,     since there is no dissipation.  The jump in $\tau$ near $z/H_p = 1$ is     caused by the forcing. \label{fig:noshear}}
\end{figure}

Equation~(\ref{eq:TGE}) can be thought of as a driven harmonic oscillator equation.  When $m$ is real and constant, its solution is a simple sinusoid. Since the transformation, equation~(\ref{eq:newvar}), compensates for the exponential decay of density with height, $w$ grows (decays) rapidly with height (depth).  Here, we have dropped the tilde overscript.  From hereon we drop all tilde overscripts and ``1'' subscripts for notational clarity.  The growth can be clearly seen in Figure~\ref{fig:noshear}, along with the corresponding constant stress (vertical transport of horizontal momentum) when there is no dissipation (see~\S\ref{sec:saturation}).

If the temperature or the flow varies with height, $m$ is a function of height $z$.  If $m$ varies slowly, the WKB approximation \citep{Bender1999} can then be used to obtain
\begin{equation}\label{eq:WKBsoln} 
  w(z)\ \approx\ 
  \frac{A\, e^{\, z/2H_{\rho}}}{m^{1/2}}                             \exp\left\{\pm\, i\! \int_{z_{\rm b}}^z m(\xi)\, \rd \xi\right\}\, . 
\end{equation}
Here, $A = w(z_{\rm b})\,[m(z_{\rm b})]^{1/2}$.  As in the constant $m$ case, the vertical perturbation velocity is wave-like with upwardly and downwardly propagating components; the amplitude of the upward component grows with height and the downward component decays with depth.  However, when $m$ varies rapidly, the solution must be obtained numerically.

At the boundaries we use the radiation condition, selecting the upwardly propagating solution at the top boundary $z_t$ and the downwardly propagating solution at the lower boundary $z_b$.  This is achieved using the condition,
\begin{equation}\label{eq:BC}
  \frac{\rd \hat{w}}{\rd z} - \left(ism + \dfrac{1}{2m} \dfrac{\rd m}{\rd z} \right) \hat{w} \ =\ 0.
\end{equation}
Here, $s=\pm 1$, depending on the signs of the horizontal and intrinsic phase speeds and on whether the condition is at the upper or lower boundary.  Note that condition~(\ref{eq:BC}) requires the WKB approximation to be valid at the boundaries.  For example, the boundaries cannot be critical layers, regions where $(c - u_0)\rightarrow 0$, since $m \rightarrow\infty$ approaching such a layer and the WKB approximation ceases to be valid.  Critical layers will be discussed further below.

As alluded to above, we use Gaussian elimination \citep{Canuto2006} in this work to numerically solve the TGE.  In the cases presented here, 3000 equally spaced levels are employed to solve for $w$.  We have checked that this resolution is adequate by performing calculations with 10,\! 000 levels and verifying convergence at the higher resolution. Extensive validations against known analytic solutions, where they exist, have also been performed.

\subsection{Polarization Relations and Fluxes}\label{subsec:pol}

The solution to the TGE is a wave in the vertical velocity perturbation.  This can be related to the horizontal and temperature (sensible and potential) perturbations.  Understanding these are essential for parameterizing the saturation process in the full non-linear situation and in general circulation models.  The perturbation quantities can be obtained from the polarization equations \citep{Hines1960}:
\begin{subequations}\label{eq:polar}
  \begin{eqnarray} 
    u\ & = &\ \frac{i}{k} \left(\frac{\rd w}{\rd z} -                   \frac{w}{H_{\rho}}   \right) \\
    \Phi\ & = &\ \frac{i}{k}\left[ \left( c - u_0 \right)                   \left(\frac{\rd w}{\rd z} -\frac{w}{H_{\rho}} \right) +
      \frac{\rd u_0}{\rd z} w \right] \\
    \theta\ & = &\ -\frac{i}{k}\left[\frac{\theta_0 N^2}
      {g \left(c - u_0\right)}\right] w \\
    T\ & = &\ -\frac{i}{k}\left[\frac{T_0 N^2} {g \left(c -           u_0\right)}\right] w\, ,
\end{eqnarray}
\end{subequations}
where we have introduced the geopotential perturbation function $\Phi$ ($\equiv p/\rho_0$).  Equation~(\ref{eq:polar}a) gives the relationship between the vertical and horizontal velocity perturbations.  Note that the amplitude of $u$ is larger than the amplitude of $w$, as it is scaled by $kH_\rho$.  Note also that the geopotential (pressure) perturbation varies with the background flow via the dependence on the intrinsic phase speed, according to equation~(\ref{eq:polar}b).  Equations~(\ref{eq:polar}c) and (\ref{eq:polar}d) shows that the potential and sensible temperature perturbations, respectively, are both $\pi/2$ out of phase with the vertical velocity perturbation.  But, the phase between $u$ and $w$ varies locally through their dependence on the background.

Gravity waves are an efficient means of transporting both momentum and energy.  The (perturbation) momentum and energy fluxes are simply obtained from equations~(\ref{eq:polar}):
\begin{subequations}\label{eq:flux}
\begin{eqnarray}
  \tau\ & = &\ \rho_0\,\overline{uw} \\
  F_x\ & = &\ \rho_0\,\overline{\Phi u} \\
  F_z\ & = &\ \rho_0\,\overline{\Phi w} \, .
\end{eqnarray}
\end{subequations}
In equations~(\ref{eq:flux}), $\tau$ is the vertical flux of horizontal momentum (or, the wave stress); $F_x$ and $F_z$ are the horizontal and vertical fluxes of energy, respectively; and, the overline indicates an average over a wavelength,
\begin{equation} 
  \overline{\alpha \beta}\ =\ 
  \frac{1}{2}\,{\rm Re}\left\{ \alpha \beta^* \right\}\, ,
\end{equation}
where $\alpha$ and $\beta$ are arbitrary complex functions and ``*'' denotes the complex conjugate.  Note that the energy fluxes depend on the background flow through their dependence on $\rho_0$ and $\Phi$.  However, as can be seen in Figures~\ref{fig:noshear} and \ref{fig:excritlayer}, the wave stress remains constant, away from the forcing and damping regions---e.g., saturation regions and critical layers.  This is in accordance with the second Eliassen--Palm theorem \citep{Eliassen1960}, which expresses non-interaction of the disturbance in the absence of dissipation and forcing.
 
In Figure~\ref{fig:excritlayer}, it is important to note that, where the fluxes are changing, the wave is interacting with the background flow in those regions.  This should be contrasted with the behavior illustrated in Figure~\ref{fig:noshear}, where the stress is not changing.  Changes in the wave stress cause accelerations to the mean flow.  Correspondingly, changes in the energy fluxes cause the temperature of the region to change.  The rates of these changes are given by \begin{subequations}\label{eq:rate} \begin{eqnarray}
  \frac{\upartial u_0}{\upartial t}\ & = & \ -\frac{1}{\rho_0} \frac{\upartial \tau}{\upartial z} \\
  \frac{\upartial T_0}{\upartial t}\ & = & \ -\frac{1}{\rho_0 c_p} \frac{\upartial F_z}{\partial z}\, . 
\end{eqnarray}
\end{subequations}
In this case, the wave causes the background flow to accelerate from 350~m~s$^{-1}$ towards 500~m~s$^{-1}$.

In the remainder of the paper, when speaking of vertically propagating waves, we consider only waves that propagate upwards from the region of excitation.  However, it must be remembered that downward propagating waves are also generated.  On a giant planet without a solid surface, those waves may not be reflected or absorbed.  They can continue to penetrate downward until they encounter a critical level or a convective region.  Or, they are simply dissipated since the amplitudes of the downwardly propagating waves decrease exponentially, as already discussed (and as also can be seen in Figure~\ref{fig:excritlayer}).  The energy and momentum fluxes are linked by the first Eliassen--Palm theorem \citep{Eliassen1960}:
\begin{equation} \label{eq:ep1}
F_z = \left(c - u_0 \right) \tau,
\end{equation}
which can be derived from equations~(\ref{eq:polar}) and (\ref{eq:flux}).  For downwardly propagating waves, we have $F_z < 0$; and, therefore, via equations~(\ref{eq:ep1})~and~(\ref{eq:rate}a), we see that for these waves the deposition of momentum also leads to acceleration of the flow toward the phase speed of the wave.  Downwardly propagating planetary scale gravity waves (i.e., thermally excited tides) are considered by \citet{Gu2009}.

\subsection{Saturation and Critical Layers}\label{sec:saturation}

\begin{figure}
  \epsscale{1.1} 
  \plotone{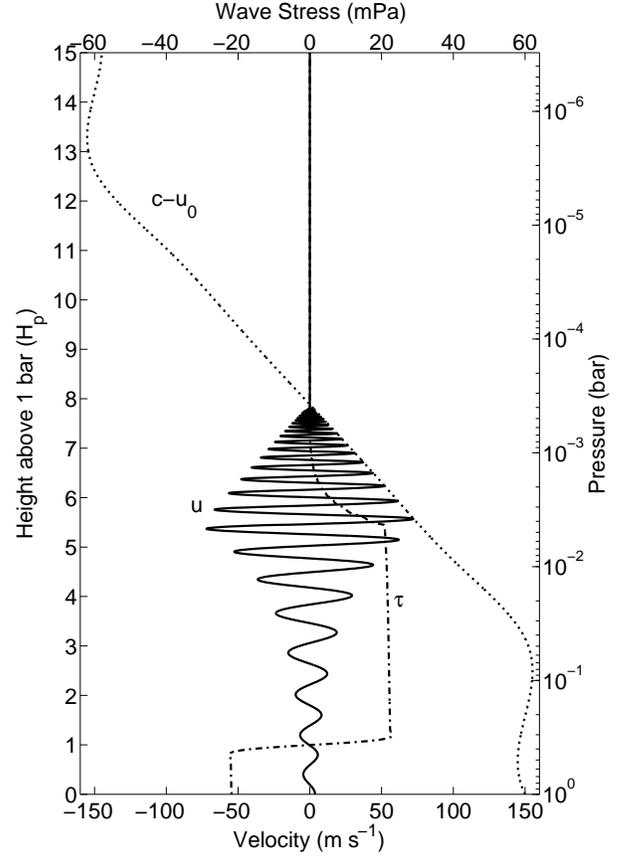}
  \caption{As in Figure~\ref{fig:noshear}, but with negative 
   vertical wind shear: $u_0$ varies from 350~m~s$^{-1}$ near 
  the bottom to 650~m~s$^{-1}$ near the top, with linear growth in between.  Here, $c = 500$~m~s$^{-1}$; hence, $(c - u_0) \rightarrow 0$ at $z/H_p \approx 8$, where the wave encounters a critical layer and dissipates.    In the region just below the critical layer, the     wave saturates and transmits momentum into the background flow, as indicated by the drop in $\tau$ with height there.       \label{fig:excritlayer}}
\end{figure}

The present work concerns inviscid, linear, monochromatic waves.  Such waves are infinite in extent and, in principle, can grow without limit when they propagate upward.  This is obviously not physical.  In reality, such waves become unstable and saturate.  The saturation process can be treated by introducing a correction to the solution according to the physical criterion,
\begin{equation} \label{eq:satcrit1} 
\frac{\upartial }{\upartial z} \left(\theta_0 + \theta \right)\ \leq \ 0,
\end{equation}
ensuring that a wave does not become convectively unstable and remains at neutral stability.  When the saturation region is identified, $\theta$ is adjusted so that neutral stability is maintained.  This new value for $\theta$ is then used in equation~(\ref{eq:polar}c) to obtain the corresponding $w$ in the saturated wave.  This value of $w$ is then used in equations~(\ref{eq:polar}a, b, c), (\ref{eq:flux}) and (\ref{eq:rate}).  This saturation condition acquires the simple form,
\begin{equation} \label{eq:satcrit2}
  \left| u \right| \ = \ \left| c - u_0 \right|,
\end{equation}
in situations where the WKB approximation is valid \citep{Fritts1984}.

As can be seen from the definitions of $N$, $H_p$, and $\theta$, the \BV\ frequency can be written as
\begin{equation} \label{eq:BVF} 
N \left(z\right) = \left[ \frac{g}{T} \left(\frac{\upartial T}{\upartial z} + \frac{g}{c_p} \right)   \right]^{1/2}.
\end{equation}
Since $g$ and $c_p$ are essentially constants in the modeled height range, $N$ depends only on the temperature profile $T(z)$.  For isothermal regions, $N$ is a constant.  In general, the fractional change of $T$ with height is small compared to $g/(Tc_p)$ throughout the modeled region.  Hence, $N$ is nearly constant in the entire domain, with a value that is roughly $2.4\!\times\! 10^{-3}$~s$^{-1}$.  The \BV \ frequency for our atmospheric profile is shown in Figure~\ref{fig:flowtempprof}.  Further discussion of the effects of variation in $N$ are presented in \S\ref{subsubsec:background}.

If the background flow contains shear, it is possible for the wave to encounter a critical layer at some height.  At the critical level, where $c = u_0$, the TGE becomes singular.  However, the equation can be solved using the method of Frobenius \citep{Bender1999}, from which it is seen that the wave is drastically attenuated by the critical layer \citep{Booker1967}.  The amount of attenuation depends on the Richardson number $Ri$ of the flow,
\begin{equation} 
  Ri\ =\ \frac{N^2}{\left(\rd u_0/\rd z\right)^2}\ . 
\end{equation}

Figure~\ref{fig:excritlayer} illustrates a critical layer encountered by a wave. If the wave stress has magnitude $\tau$ below the critical layer, it then has magnitude, $\tau \cdot \exp \left\{-2 \pi \left[Ri - (1/4)\right]^{1/2} \right\}$, after the encounter with the critical layer \citep{Booker1967}.  This is a substantial amount of attenuation.  For example, in our model atmosphere of Figure~\ref{fig:excritlayer}, $Ri \ga 900$.  Hence, the wave is essentially completely dissipated at the critical level, with the wave stress falling to practically zero and the momentum deposited into the mean flow.  Note that, during its approach to the critical layer, the wave actually saturates and deposits momentum and energy over a finite layer (cf., Figure~\ref{fig:excritlayer}).  It is important to note that while a wave saturates when approaching a critical layer, the presence of a critical layer is {\it not} required for saturation.  Saturation is a general process referring to wave dissipation by many different mechanisms---e.g., radiation, conduction, breaking, and turbulence.

Numerically, when a critical layer is present, we lift the phase speed from the real axis by adding a small imaginary component: $c = c_r + ic_i$, where $|c_i / c_r| < 10^{-3}$.  This introduces a small amount of linear damping and ensures that the neglected nonlinear terms do not dominate in the regions where waves become steep and eventually break.  Of course, adding damping causes the wave-stress to decrease with height and the second Eliassen-Palm theorem to be no longer valid.  However, the effect is small.  This can be seen in Figure~\ref{fig:excritlayer}, where the wave stress falls negligibly over the layer from $z/H_p \approx 1$ to $z/H_p \approx 5$.

\section{Extrasolar Planet Application}\label{sec:application}

\subsection{Setup}

Our aim in the present paper is to demonstrate several properties of gravity waves likely to be important for hot extrasolar planets.  For this, we choose \HD\ as a paradigm planet. This planet has been the focus of many theoretical and observational studies, and it is expected to be generic with respect to the properties discussed here.  The physical parameters that characterize the planet's atmosphere are given in Table \ref{tab:params}.

\begin{table}
\begin{center}
  \caption{Parameters Used for ${\rm HD}\,209458\,{\rm b}$}
\begin{tabular}{llr}
  \hline
  \\
  Specific Gas Constant & $R$ & 3523~J~kg$^{-1}$~K$^{-1}$ \\
  Specific Heat at Constant Pressure \hspace*{5mm} & $c_p$             \hspace*{3mm} &   12300~J~kg$^{-1}$~K$^{-1}$ \\
  Acceleration Due to Gravity & $g$ & 10~m~s$^{-2}$ \\
  Rotation Rate & $\Omega$ & 2.08$\times 10^{-5}$~s$^{-1}$ \\
  \\
  \hline
\end{tabular}\label{tab:params}
\end{center}
\end{table}

\subsubsection{Forcing}

In a stratified atmosphere, gravity waves are readily generated by many mechanisms---thermal and mechanical---such as the absorption of stellar radiation, convective release of latent heat, storms, flows over topography and coherent localized ``heat islands'', and detonation by impacts. In this work, we consider small- and meso-scale thermally excited waves, rather than large-scale waves, as already discussed.  The horizontal wavelengths used in this work are 2500~km or less.  This is a reasonable range since it is well within the observed range of gravity waves on Jupiter \citep{Young1997,Hickey2000}.  Although not unimportant, we do not dwell on the precise nature of the source of the excitations. The main interest here is in the propagation and deposition of momentum and energy.

The forcing in equation~(\ref{eq:TGE}) is simply represented as a Gaussian, modified so that it is zero beyond two half-widths above and below the center. The center is located at $z/H_p = 1$ above the bottom of the domain.  The half-width is 75~km, or $\sim0.15 H_p$.  The forcing location and width are chosen so that the vertical scale of the forcing is less than the vertical wavelength of the waves we present here.  We have extensively explored the effects of varying the location, width, and strength of forcing and present this case to illustrate several important points.  Not surprisingly, the dynamics do depend on the chosen parameter values, but the dependence is broadly predictable.  For example, if the forcing scale is much larger than the vertical wavelength, then only a very small amplitude wave is emitted from the forcing region, due to cancellations.

The heating rate, $\dot{Q} / c_p$, is set to 10$^{-3}$~K~s$^{-1}$. Note that this is a modest value.  The forcing corresponds to roughly 300~K per rotation of the planet.  This is compared to $\sim$100~K per rotation at the chosen location in the circulation model of \citet{Thrastarson2009}.  A forcing of $\sim$1100~K per rotation, for a similar latitude-longitude location on the planet (see \S\ref{subsubsec:background}), is used in \citet{Showman2008}. The latter value implies that, in the absence of motion, the location on the planet will cool completely in one rotation of the planet.  We stress that locally---i.e., scales far below the grid scale of the current circulation models---the forcing could actually be much stronger. The actual value is presently uncertain and likely to be spatially and temporally variable over the planet. To provide a context, for the Earth the heating rate is $\sim\! 2$~K per day (1 day = 0.29 rotation of \HD) over large areas; but, locally, at the tops of low clouds on the Earth the rate can be up to $\sim\! 50$~K per day \citep{Wallace2006}.

\subsubsection{Background Structure}\label{subsubsec:background}

\begin{figure}
  \vspace{2pt}
  \epsscale{1.1}
  \plotone{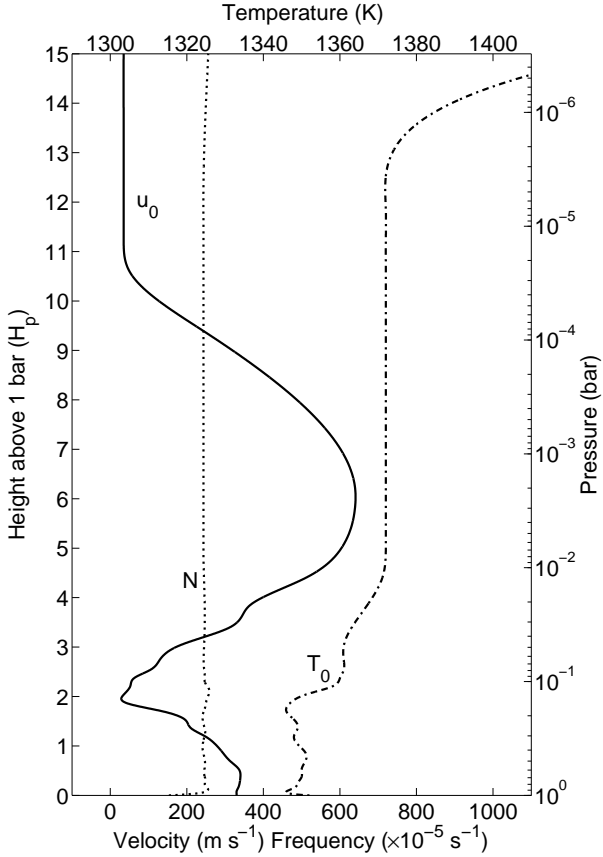}
  \caption{Sample atmospheric mean flow $u_0$ (---), temperature $T_0$     ($-\!  \cdot\!  -$), and \BV \ frequency, $N$ ($\cdots$) profiles     of a typical hot extrasolar planet \HD, used in this work.  The     profile is representative of a region at approximately     70$^{\circ}$E,\,10$^{\circ}$N.  The     planet is a close-in giant planet.  The profiles are obtained from     a 3-D, global circulation model, up to $\sim\! 10^{-3}$~bar level.     Above that level the profiles are simply extended, following     loosely the observed profiles of Jupiter     \citep{Young2005,Flasar2004}.  \label{fig:flowtempprof}}
\end{figure}

Figure~\ref{fig:flowtempprof} shows the mean flow and temperature profiles used to obtain much of the results presented in this section. The lower part of both profiles---approximately the lower six scale heights---is taken from global circulation simulations of the hot extrasolar giant planet \HD\ by \citet{Thrastarson2009}, using the NCAR Community Atmosphere Model \citep{Collins2004}.  The profile is from a point near the equator, slightly away from the substellar point (70$^{\circ}$E,\,10$^{\circ}$N). This point was chosen as it is within the equatorial jet and the Coriolis parameter $f= 2 \Omega \sin \phi$ is not large. The temperature profile generally increases with height over the lowest 4 scale heights and then becomes isothermal. Fortuitously, this provides an opportunity for a loose validation of our model: it is very similar to the temperature structure observed by the Galileo probe in the same region of Jupiter's atmosphere \citep{Young2005}.  We extend the profile by keeping the temperature isothermal through the planet's stratosphere and having the thermosphere (beginning of the temperature inversion near the top) start between 12 and 13 scale heights at $p_0 \approx 4 \times 10^{-6}$~bar.  This profile has a Richardson number of at least 3.4, giving an attenuation of $1.4 \times 10^{-5}$.  Hence, any critical layer can be considered to fully dissipate the wave.

\begin{figure}
  \epsscale{1.08} 
  \plotone{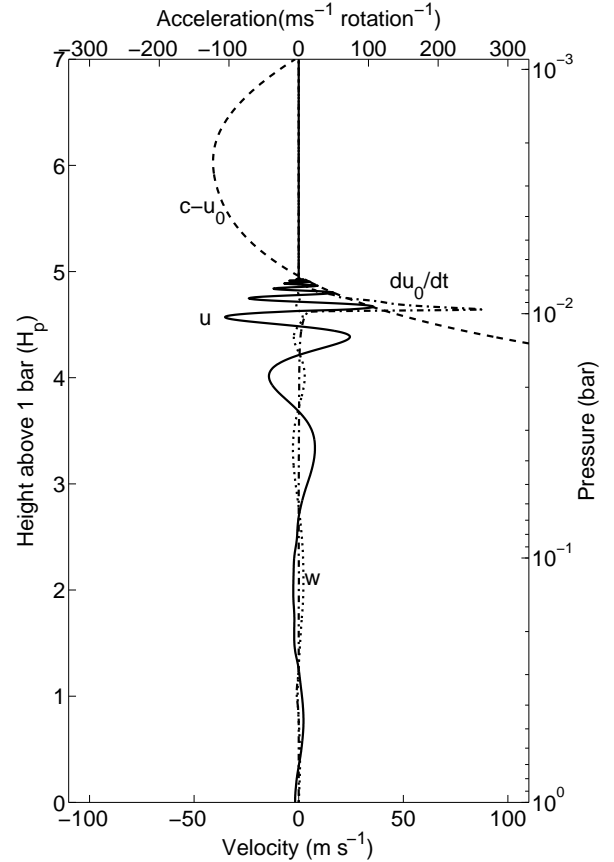}  
  \caption{A gravity wave with $c = 600$~m~s$^{-1}$, propagating in an     atmosphere with profiles shown in Figure~\ref{fig:flowtempprof}.  The     horizontal perturbation velocity $u$ (---), intrinsic phase speed, $c -     u_0$ ($\cdots$), and the mean flow acceleration $\rd u_0/\rd t$ ($-\!     \cdot\! -$) are shown.  The intrinsic phase speed becomes zero at $z/H_p     \approx 5$ and the wave encounters a critical level.  In the layers just     below the critical level the wave saturates and sheds momentum into the     mean flow, causing it to accelerate, peaking at a rate over     250~m~s$^{-1}$~rotation$^{-1}$. \label{fig:realcritlayer}}
\end{figure}

The chosen flow profile has two local flow velocity maxima. The upper maximum is extended into a jet with a peak at $z/H_p \approx 6$. This is similar to the structure of the jet in Jupiter's stratosphere proposed by \citet{Flasar2004}, which has a peak velocity between the 10$^{-2}$ and 10$^{-3}$~bar levels.  We then extend the profile further upwards without shear.  Note, this is in keeping with the lower boundary of the model of \citet{Koskinen2007}.  The structure is somewhat different than those of \citet{Showman2008,Showman2009}, where there is just one jet with the peak approximately located at the 10$^{-1}$~bar level.  The peak flow speed is also much greater in those studies at 4 or 5~km~s$^{-1}$.  It is important to note, however, that these differences do not change qualitatively the basic points we are making in this paper.

The \BV\ frequency profile $N(z)$ is also shown in Figure~\ref{fig:flowtempprof}.  As already discussed, the profile does not vary much over the whole domain: the maximum \BV\ frequency is just 1.2 times the minimum value.  Therefore, $N$ does not contribute much to the variation of the index of diffraction $m$.  The main contributor to the variation of $m$ is the variation of the intrinsic phase speed, which is derived from the large variation of flow speeds.  This should be compared to the analogous terrestrial situation, where the range of flow speeds is much lower.  This allows $N$ to have a larger effect on the variation of $m$ on the Earth.

\subsection{Wave-Background Interaction}\label{subsec:WBI}

\subsubsection{Critical Layer Encounter}

Figure~\ref{fig:realcritlayer} shows a gravity wave encountering a critical level in the upper jet in Figure~\ref{fig:flowtempprof}.  The wave has a horizontal wavelength, $2\pi/k$, of 2500~km and $c = 600$~m~s$^{-1}$. Here, since $c > u_0$ as the wave approaches the critical layer from below, the momentum deposited in the mean flow causes the mean flow to accelerate.  This acceleration peaks at over 250~m~s$^{-1}$ per rotation.  This is large enough to double the flow speed at this layer in $\sim$2~rotations---a significant effect.  The effect is large enough to require its inclusion in any simulation of the atmospheric circulation \citep{Cho2008a}.

The waves encountering critical layers are dissipated.  Therefore, a flow with a range of flow speeds dissipates all gravity waves with phase speeds within this range.  That is, a spectrum of gravity waves is prevented from propagating high into the atmosphere.  There are other, secondary effects at the critical layer that may also affect the mean flow; but, they are not modeled here.  They will form the basis of future work.  For example, the deposition of energy into the flow at the critical layer may well lead to the generation of new gravity waves, which then may propagate further, partly mitigating the filtering effect.

\begin{figure}
  \epsscale{1.1} 
  \plotone{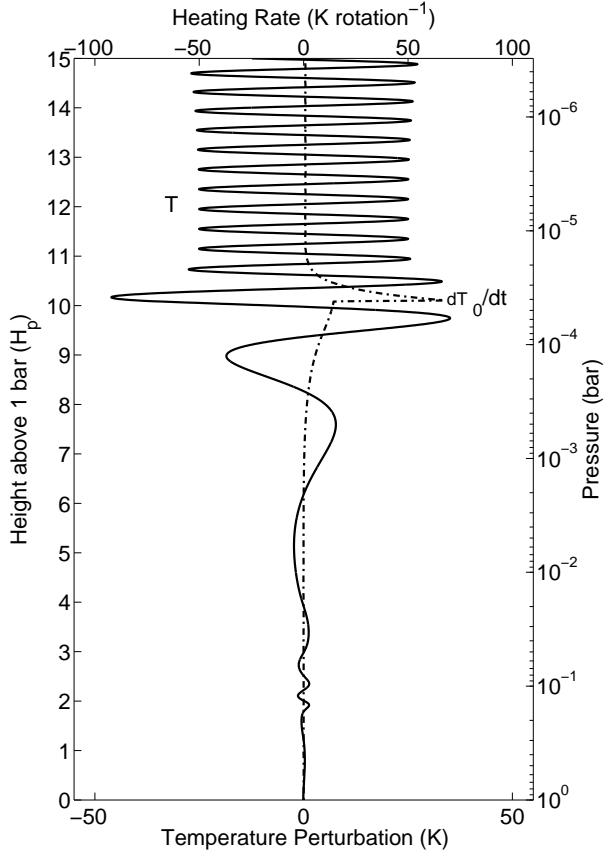}
  \caption{A gravity wave, with $c = -40$~m~s$^{-1}$, propagating in     an atmosphere described by the profiles in     Figure~\ref{fig:flowtempprof}.  The perturbation to the     temperature field $T$ (---) and the heating rate $\rd T_0 / \rd t$     ($-\,\cdot\, -$) are shown.  The wave saturates at just above     $z/H_p = $~10, where the heating peaks at nearly 80~K~rotation$^{-1}$. The peak energy flux for this wave is approximately 1~W~m$^{-2}$.} \label{fig:realsaturate}
\end{figure}

\begin{figure}
  \vspace{2.5pt}
  \epsscale{1.1}
  \plotone{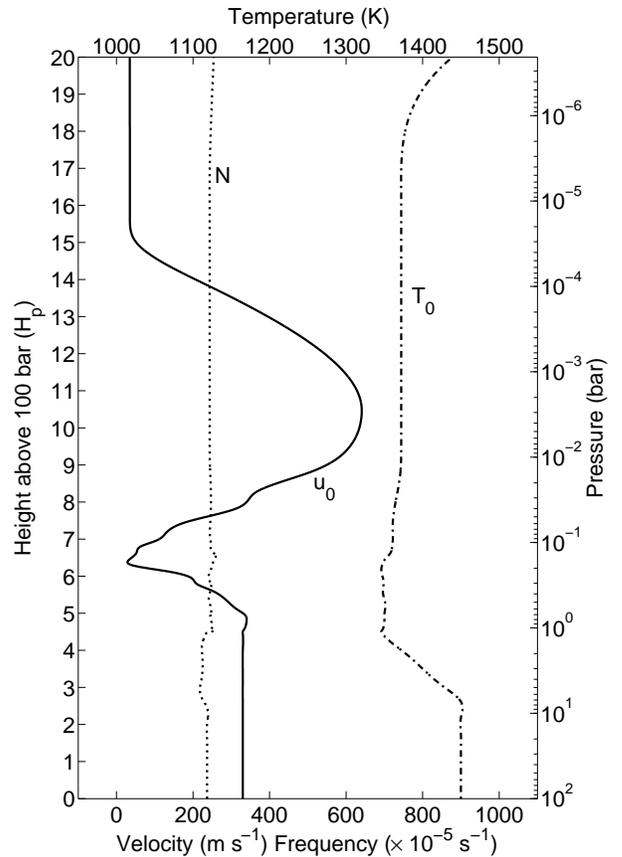}
  \caption{As in Figure~\ref{fig:flowtempprof}, but with the bottom of     the computational domain extended down to 100~bars.  Here, $u_0$     is extended downward barotropically (independent of height) from     the 1~bar level; $T_0$ is extended downward so that the profile     below the 1~bar level is similar to that of Figure~18 in     \citet{Showman2009}.  \label{fig:flowtempprofdeep}}
\end{figure}

\subsubsection{Saturation}

Figure~\ref{fig:realsaturate} shows an example of a gravity wave saturating in the atmosphere of Figure~\ref{fig:flowtempprof}.  In general, it is possible that a wave may not encounter a critical level as it propagates upward.  However, such a wave then travels higher into the atmosphere, where it grows large and, if unabated, eventually breaks or suffers dissipation at higher altitudes.  Although both momentum and energy are deposited in this case, we focus here on the effects on the temperature field.

The wave launched in Figure~\ref{fig:realsaturate} has $c=-40$~m~s$^{-1}$ (i.e., westward).  The horizontal wavelength remains at 2500~km, as in the critical layer example of Figure~\ref{fig:realcritlayer}.  The vertical velocity perturbation grows with height.  Therefore, so do the zonal velocity perturbation $u$ and the potential temperature perturbation $\theta$, as expected from equation~(\ref{eq:polar}).  In this case, the wave saturates near the top of the jet, where condition~(\ref{eq:satcrit1}) is exactly satisfied.  Here, condition (\ref{eq:satcrit2}) is approximately satisfied since the zonal perturbation velocity and the intrinsic phase speed ($c - u_0$) are both approximately 122~m~s$^{-1}$.  The saturation deposits energy that causes the atmosphere there to heat up.  The heating is significant, peaking at $\sim\! 75$~K per rotation---a 5\% change in one rotation.  In the absence of other effects, the ambient temperature can be doubled in about 20 planetary rotations (or orbits, assuming 1:1 spin-orbit synchronization).

A wave that has a phase speed greater than the maximum flow speed will not encounter a critical layer.  Those with phase speeds close to, but still above, the maximum flow speed will, in general, saturate in the regions just below the maximum flow, as the intrinsic flow speed will be small in that region. Similarly, waves with phase speeds just less than the minimum flow will saturate as well.  In this way the filtering effect discussed above is extended beyond those waves with phase speeds equal to flow speeds.  The main effect of these filtered waves on the flow will be lower in the atmosphere. In the profile given in Figure \ref{fig:flowtempprof}, where the waves emanate from the $z/H_p = 1$ level, this means that the upper layers of the lower jet will be slowed by gravity waves whereas the lower levels of the upper jet will be accelerated.

Those waves that do not dissipate will be able to propagate into the upper atmosphere depositing their momentum and heat there. Here the changes to the flow can be very large.  For example, the wave shown in Figure~\ref{fig:realsaturate} causes a deceleration of up to 6.8~km~s$^{-1}$~rotation$^{-1}$ as it saturates.  This clearly dominates the flow at this level.

Moving the location of the wave origin down does not change the basic behavior in the qualitative sense.  However, the amplitudes are much larger, compared with the case when the wave originates higher up in altitude.  Thus, the possibility exists for stronger effects due to gravity wave interaction with the background.

\begin{figure}
  \vspace{3pt}
  \epsscale{1.1}
  \plotone{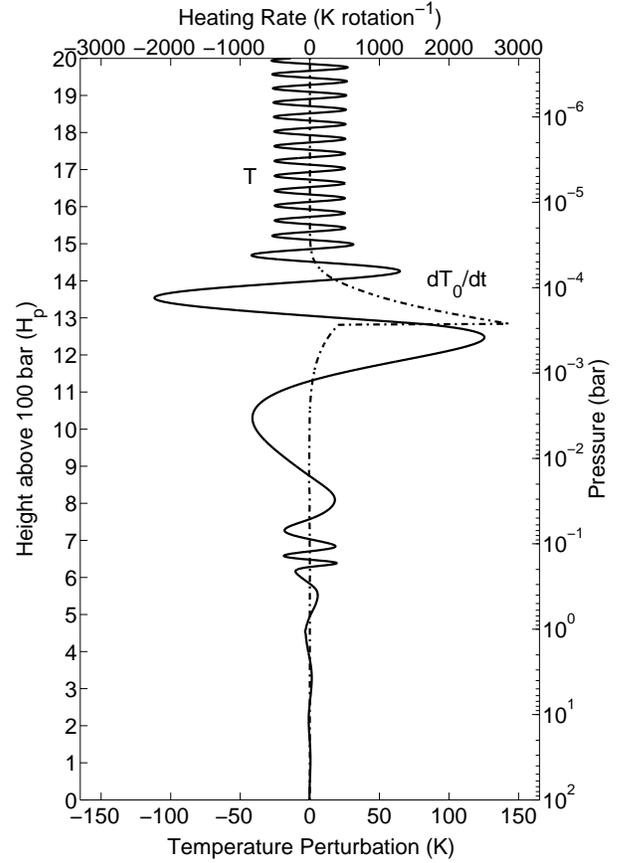}
  \caption{A gravity wave, with $c = -40$~m~s$^{-1}$, propagating in     an atmosphere described by the profiles in     Figure~\ref{fig:flowtempprofdeep}.  The perturbation to the     temperature field $T$ (---) and the heating rate $\rd T_0 / \rd t$     ($-\,\cdot\, -$) are shown.  The wave saturates at just above     $z/H_p = 13$, where the heating rate peaks at just under     3000~K per rotation.  In terms of the pressure level, this     location is actually lower than in the case shown in     Figure~\ref{fig:realsaturate}, and the magnitude of the peak is     nearly 50~times greater. The peak energy flux for this wave is nearly 200~W~m$^{-1}$. Thus, having a source at a lower height     can have a much stronger effect.  \label{fig:realsaturatedeep}}
\end{figure}

This is illustrated in Figures~\ref{fig:flowtempprofdeep} and \ref{fig:realsaturatedeep}, which should be compared with Figures~\ref{fig:flowtempprof} and \ref{fig:realsaturate}. Here, we have extended the profiles downwards.  The wave is still launched at $z/H_p = 1$, but this is now deeper in the atmosphere.  The wave again has a phase speed of $-40$~m~s$^{-1}$, and the horizontal wavelength remains at 2500~km.  This wave also saturates near the top of the upper jet, where the energy deposition into the mean flow causes the atmosphere there to heat up.  Note that the region of heating is lower than when the wave originates higher up, as in Figure \ref{fig:realsaturatedeep}.  The heating is significant, peaking at $\sim\! 3000$~K~rotation$^{-1}$.  The ambient temperature can be doubled in approximately half of a planetary rotation.  In a more realistic scenario, dissipation---which we have not included in our model---will likely reduce the heating rate.
 
\begin{figure}
  \epsscale{1.08}
  \plotone{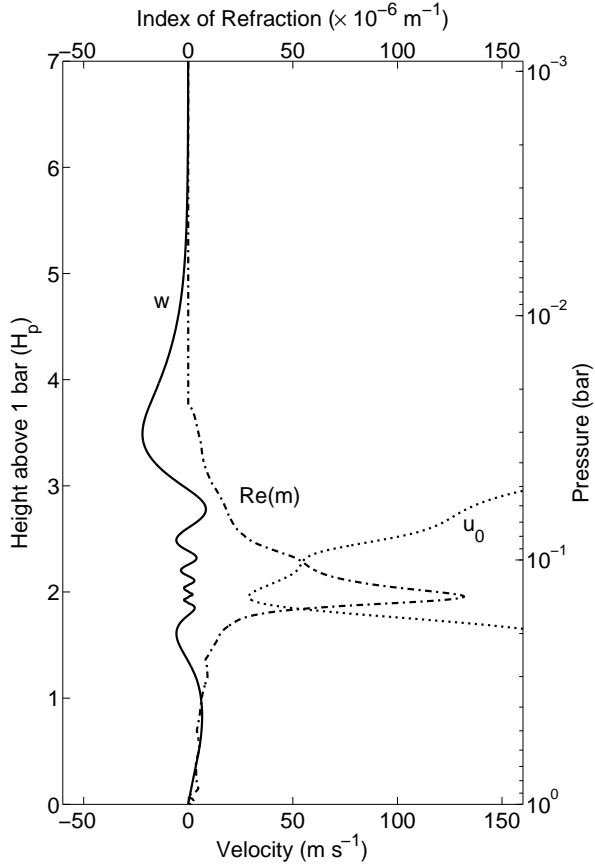}
  \caption{A gravity wave, with $c = 10$~m~s$^{-1}$ and horizontal     wavelength $2\pi / k = 955$~km, trapped in an atmosphere with the     structure presented in Figure~\ref{fig:flowtempprof}.  The     vertical velocity perturbation $w$ (---), mean flow $u_0$ ($-\,     -$), and the real part of the index of refraction $m$ ($-\cdot -$)     are shown.  The wave is trapped in relatively quiescent region between $z/H_p     \approx 1.5$ and $z/H_p \approx 3.5$ and does not propagate     vertically.  The region of trapping corresponds to the region     where $m$ is real.  The wave is reflected at the boundaries of     this region, providing a possibility for     resonance. \label{fig:realtrapped}}
\end{figure} 
 
\subsubsection{Refraction}\label{subsubsec:refract}

So far we have been focusing on the vertical transport of momentum and energy by gravity waves.  However, the waves can also transport momentum and energy {\it horizontally} (cf., \S\ref{subsubsec:ducting}).  Substituting $c = \omega / k$ into the the index of refraction, equation~(\ref{eq:indref}), and rearranging gives the dispersion relation for gravity waves.  Here, we consider the case where $u_0 = 0$.  We will examine cases where $u_0 \ne 0$ in the sections following this one.

When there is no background mean flow, the dispersion relation simplifies to
\begin{equation} \label{eq:dispersion} 
\omega\ =\ \pm\frac{Nk}{\left[k^2 + m^2       + 1/(4H_{\rho}^2)\right]^{1/2}}.
\end{equation}
We can then use the definitions,
\begin{subequations} \label{eq:grpvel}
\begin{eqnarray} 
  u_g &\ =\ & \frac{\upartial \omega}{\upartial k} \\
  w_g &\ =\ & \frac{\upartial \omega} {\upartial m},
\end{eqnarray}
\end{subequations}
to obtain the group velocities.  They are:
\begin{subequations} \label{eq:noflowgrpvel}
\begin{eqnarray} 
  u_g &\ =\ & \pm\frac{N\left[m^2 + 1/(4H^2_{\rho})\right]}{\left[k^2 + m^2 + 1/(4H^2_{\rho})\right]^{3/2}} \\
  w_g &\ =\ & \pm\frac{N k m}{\left[k^2 + m^2 + 1/(4H^2_{\rho})\right]^{3/2}}.
\end{eqnarray}
\end{subequations}
Thus, for propagating waves (i.e., waves for which $m$ is real), $u_g \neq 0$. Therefore, gravity waves always propagate obliquely and cannot strictly propagate vertically when there is no background flow.

From equations~(\ref{eq:noflowgrpvel}) we see that $\vartheta$, the angle of propagation with respect to the horizontal, is given by
\begin{equation}
\tan \vartheta\ =\ \frac{k m}{m^2 + 1/(4H^2_{\rho})}.
\end{equation}
Since $H_{\rho}$ is nearly constant, with a value just under 500~km, $\tan \vartheta$ varies with $1/m$ for $m\gtrsim10^{-6}$~m$^{-1}$.  This gives rise to refraction.  As a wave propagates into a region of higher $m$, it bends to a more horizontal path.  Note that, as we are here considering a region with no flow, increasing $m$ is essentially equivalent to increasing $N$.  As shown in Figure~\ref{fig:flowtempprof}, $N$ increases in the thermosphere and so the paths of waves in this region will bend towards the horizontal even though the flow is small.

\subsubsection{Trapping}\label{trapping}

From equation~(\ref{eq:WKBsoln}) we can see that in regions where $m$ is imaginary, the wave is evanescent: its amplitude decays towards zero and therefore it does not propagate vertically.  This can occur when $N^2 < 0 $ (i.e., when the atmosphere is convectively unstable).  But, it can also occur for non-hydrostatic waves when the buoyancy term becomes dominated by the non-hydrostatic term.  In addition, when the index of refraction changes between layers, the wave is reflected at the boundary.  The amount of reflection is given by the magnitude of the coefficient of reflection $|r|$, where 
\begin{equation}\label{eq:coeff_reflect}
  r\ =\ \frac{m_1 - m_2}{m_1 + m_2}\, .
\end{equation}
In equation~(\ref{eq:coeff_reflect}), $m_1$ and $m_2$ are the indices of refraction in two adjacent layers.  When $m_2$ is imaginary, total reflection occurs and the wave is evanescent in that region and its amplitude decays to zero.  However a region of propagation can exist between two regions of evanescence.  This readily occurs for jets, where the intrinsic phase speed varies enough to allow the hydrostatic term to dominate in some regions and not in others.  The region can also occur through variations of the \BV\ frequency.  In Figure~\ref{fig:realtrapped} we see a wave that is trapped in the relatively quiescent region between $z/H_p \approx 1$ and $z/H_p \approx 4$.  Outside this region, the value of Re$(m)$ is small, or zero.  Trapped in this region the wave is able to interact with itself and, under appropriate conditions, resonate.

This is another mechanism via which waves may be filtered out by the flow; and so the waves do not reach high altitudes at which saturation can occur.  However, in this case, a trapped wave does not directly interact with a low level flow that changes its characteristics.  Indeed, between the two reflection layers the wave can propagate horizontally---even in the absence of any flow, using the refractive mechanism described above.  As long as the layers do not allow much leakage, it is possible for a trapped wave to cover large horizontal distances---transporting momentum and heat zonally (east-west direction).

\subsubsection{Ducting}\label{subsubsec:ducting}

\begin{figure}[t]
  \epsscale{1.1}
  \plotone{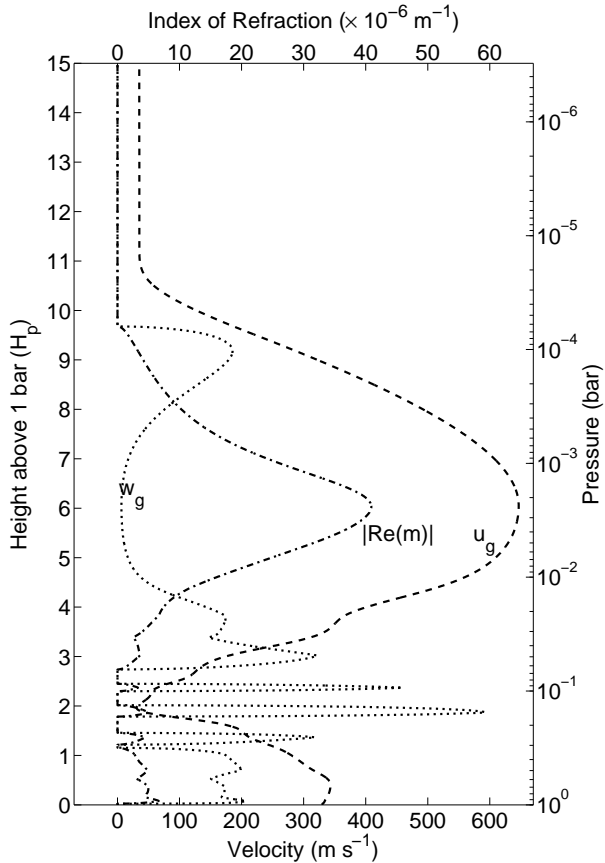}
  \caption{As in Figure~\ref{fig:realtrapped}, but with $c = 700$~m~s$^{-1}$ and horizontal wavelength $2\pi / k = 1410$~km.  
The horizontal group speed $u_g$ ($-\, -$), vertical group speed $w_g$ ($\cdots$) and the real     part of the index of refraction $m$ ($-\cdot -$) are shown.  The     wave, not shown, is trapped in the upper jet between $z/H_p \approx 3$ and     $z/H_p \approx\! 9$, the region where $m$ is real, and does not propagate vertically above this region; it is     however, able to propagate along the jet as the large value of $u_g$ within the jet shows.  The wave is     reflected at the boundaries of this region, providing a possibility for resonance.  \label{fig:realducted}}
\end{figure}

As well as being trapped in relatively quiescent regions, it is possible for waves to be trapped in a jet.  As alluded to above, it is possible for such a wave to travel within the region of trapping, which is known as a duct or a waveguide.  In Figure~\ref{fig:realducted}, a non-hydrostatic wave with $c = 700$~m~s$^{-1}$ is trapped within the jet (located at $\sim$5~mbar level) in our model atmosphere.  Note the small values of Re$(m)$ outside the jet.

In this case, the flows are significant. Therefore, we use the full dispersion relation,\\
\begin{widetext} 
\begin{equation} \label{eq:fulldispersion}
\omega\ =\ \frac{ku_0 + 2H_{\rho}k\left[u_0'+H_{\rho}\left(2\left(k^2+m^2\right)u_0+u_0''\right)\right] \\ \pm 2\left[H^2_{\rho}k^2\left(\left(1+4H^2_{\rho}\left(k^2+m^2\right)\right)N^2 +\left(u_0'+H_{\rho}u_0''\right)^2\right)\right]^{1/2}}{1+4H^2_{\rho}\left(k^2+m^2\right)} \, ,
\end{equation}
\end{widetext}
to develop expressions for the group velocities.

\begin{figure}
  \epsscale{1.1}
  \plotone{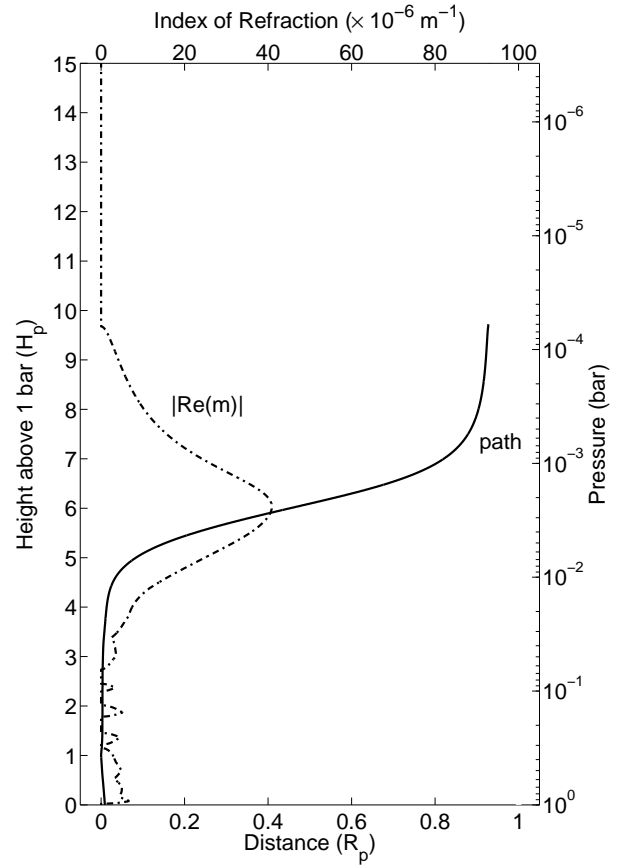}
  \caption{This shows the path of propagation of the wave in Figure~\ref{fig:realducted} assuming that properties of the duct do not change in the $x$-direction. The path shown is the first crossing of the duct, that is until the wave encounters the top reflection layer. At this point the wave will reflect and then propagate downwards in a mirror image of this path. Note that the wave travels nearly one planetary radius before reflection. This means with just six reflections the wave will have nearly circumnavigated the planet. Of course in the real situation the properties of the duct will change in the $x$-direction and the wave will probably either leak out of the duct or dissipate before the circumnavigation is complete. \label{fig:realductedpath}}
\end{figure}

However, the expressions obtained are large and rather unilluminating.  They can be simplified by assuming that $u_0'$ and $u_0''$ are small.  This is realistic since the shear is of the order of $10^{-3}$~s$^{-1}$ and $u_0''$ of the order $10^{-8}$~m$^{-1}$~s$^{-1}$.  This is small compared with the other terms in the expressions.  This then gives \begin{subequations}
  \begin{eqnarray} \label{eq:grphorvelflow}
  u_g &\ =\ & u_0+\frac{N\left[m^2 + 1/(4H^2_{\rho})\right]}{\left[k^2 + m^2 + 1/(4H^2_{\rho})\right]^{3/2}} \\
  w_g &\ =\ & \pm\frac{N k m}{\left[k^2 + m^2 + 1/(4H^2_{\rho})\right]^{3/2}}	
  \end{eqnarray}
\end{subequations}
From these expressions, we can see that $u_g$ follows $u_0$ as this is the larger term on the right hand side of equation~(\ref{eq:grphorvelflow}) in our model atmosphere.  In Figure~\ref{fig:realducted}, the values of $u_g$ and $w_g$ are shown.  Note that in the center of the duct $w_g$ is very small while $u_g$ is large, so that energy is transported along the flow.  At the the top and bottom of the duct the vertical group velocity increases, while the horizontal group velocity falls.  Therefore, propagation here is nearly vertical.  In Figure~\ref{fig:realducted}, we show $w_g$ as positive, however this is only for the upward propagation of energy, at the top of the duct the wave is reflected and the vertical component becomes negative.  This keeps the wave within the jet.  The ray path followed by the wave, before the reflection, is shown in Figure~\ref{fig:realductedpath}.

The wave can travel large distances in this duct; but, eventually, the wave will either escape the duct or be dissipated. The range of speeds in the jet may change so that a critical level for the wave is created.  The wave will then be reabsorbed into the flow.  Alternatively, if the flow or \BV\ frequency outside the jet changes so that the buoyancy term is no longer small and is dominated by the non-hydrostatic term, then the reflection is no longer total and the wave will then leak out of the duct.  This can, for example, happen when propagating into a colder region, assuming the lapse rate remains constant.  As can be seen from equation~(\ref{eq:BVF}) a fall in temperature with constant lapse rate will cause an increase in the \BV\ frequency and thus an increase in the buoyancy term. This may occur very far from the original region of wave excitation. Indeed, in the example given, it is possible to envisage jets ducting waves and so transporting energy from the dayside of a tidally locked planet to the colder nightside where the waves escape the jet and propagate away from the duct before dissipating.

\section{Implications for Circulation Models}\label{sec:implications}

The effects of gravity waves discussed in this work on the larger-scale circulation must be parameterized in global models because the spatial resolution---both horizontal and vertical---required to model them is currently prohibitive.  The waves important to the large-scale extrasolar planet atmospheric circulation have horizontal length scales ranging from approximately $\sim\!  10^5$~m to $\sim\! 10^7$~m and vertical wavelengths as small as $\sim\!  10^4$~m.  Waves with periods of few hours can carry significant momentum and energy flux vertically, but the sources of these waves include processes that are not included or resolvable in current circulation models.

The difficulty with representing gravity waves in GCMs exists even for the GCMs of the Earth.  For example, the parameterization for convection is not aimed at producing realistic gravity waves \citep{Collins2004}.  However, not representing gravity waves can affect the accuracy of the GCMs.  The lack of gravity wave drag can lead to the overestimation of wind speeds, resulting in faster and narrower jets than observed \citep{McLandress1998}.  Further, gravity waves introduce turbulence with subsequent mixing and thermal transport \citep{Fritts2003}.  This leads to greater homogenization of the atmosphere with a reduction in, for example, temperature gradients.  Gravity waves also interact with planetary waves, playing a role in important transient phenomena (such as sudden stratospheric warming).  In the absence of gravity waves, these phenomena are not accurately modeled \citep{Richter2010}.

There are many parametrization schemes currently incorporated or proposed for general circulation modeling \citep{McLandress1998}.  In all of the schemes, the basic components are 1) specification of the characteristic of the waves at the source level, 2) wave propagation and evolution as a function of altitude, and 3) effects on and by the atmosphere.  All of them are essentially linear and one-dimensional, in that waves only propagate vertically and that only vertical variation in the background influence the propagation.  As seen in this work, linear theory still requires information such as the wave's phase speed $c$ and wavenumber $\mathbi{k}$, for example. A more complete theory would need spatial and temporal spectral information. Intermittency is another crucial feature that would need to be taken into account.  The primary differences in various schemes pertain to the treatment of nonlinearity and specificity of wave dissipation mechanisms.
 
Currently, all global circulation models of hot Jupiters suggest the presence of a low number of zonal jets.  However, all the models do not have the resolution required to adequately resolve gravity waves and are subject to all of the limitations described above.  This issue has been previously raised by \citet{Cho2008a}, in which they advocate caution against quantitative interpretation of current model results. For example, without the inclusion of the wave effects discussed in this work, high jet speeds and precise eastward shift of putative ``hot spots'' can be questioned \citep[e.g.,][]{Cooper2005,Knutson2007,Langton2007} 

\section{Conclusion}\label{sec:conclusion}

Gravity wave propagation and momentum and energy deposition are complicated by the environment in which the wave propagates.  For example, spatial variability of the background winds causes the wave to be refracted, reflected, focused, and ducted. Additionally, temporal variability of the background winds cause the wave to alter its phase speed.  Still further complications arise due to the wave's ability to generate turbulence, which can modify the source or serve as a secondary source, and the wave's interaction with the vortical (rotational) mode of the atmosphere.  Many of these issues are as yet not well-understood and are currently areas of active research.

In this work, we have emphasized only some of these issues.  We have shown that gravity waves propagate and transport momentum and heat in the atmospheres of hot extrasolar planets and that the waves play an important role in the atmosphere.  They modify the circulation through exerting accelerations on the mean flow whenever the wave encounters a critical level or saturates.  They also transport heat to the upper stratosphere and thermosphere, causing significant heating in these regions.  Moreover, through ducting, they also provide a mechanism for transporting heat from the dayside of tidally synchronized planets.

Before relying on GCMs for quantitative descriptions of hot extrasolar planet atmospheric circulations, further work needs to be performed to ensure that the effects of important sub-scale phenomena, such as the gravity waves discussed here, are accurately parameterized and included in the GCMs.

\section*{Acknowledgments}
The authors thank Heidar Thrastarson for generously sharing information from his simulation work on extrasolar planet atmospheric circulation.  We also thank Orkan Umurhan for helpful discussions and the anonymous referee for helpful suggestions. C.W. is supported by the Science and Technology Facilities Council (STFC). J.Y-K.C. is supported by NASA NNG04GN82G and STFC PP/E001858/1 grants.\\

\end{document}